# The Statistical Analysis of Pairwise Experiments with Qualitative Responses

A. H. Al-Ibrahim

P.O. Box 430, Surra, 45705

Kuwait

dr.alibrahim@gmail.com

## ABSTRACT

Suppose an experiment is conducted on pairs of objects with outcome responses a continuous variable measuring the interactions among the pairs. Furthermore, assume the response variable is hard to measure numerically but easy to be coded into ordered categories such as low, moderate, and high levels of interaction. In this paper we estimate the unknown interaction values from the information contained in the coded data and the design structure of the experiment. The method of estimation is shown to enjoy several optimal properties such as explaining maximum variance in the responses with minimum number of parameters and for any probability distribution underlying the responses. Other properties of the method include: the interactions have the simple interpretation of correlation, size of error is estimable from the experiment, and only a single run of each pair is needed to carry out the experiment. We also explore possible applications of the technique. Three applications are presented, one on protein interaction, a second on drug combination, and the third on computer imaging. The first two applications are illustrated using real life data while for the third application the data are generated via binary coding of an image.

1-Introduction

Interaction is a familiar notion and is used in everyday language to indicate communication or feedback between individuals and/or objects. In science, it is often coined with statistical analysis of multi-factor experiments to imply the joint effects of two or more factors is different of the gross effects of the individual factors. In recent years, however, interaction has been used in new contexts or applications. For example, biochemists talk about protein interaction to refer to the act of physical contact of protein molecules to form sequences of proteins, and gene interaction to refer to the interplay between multiple genes that impact gene expression or phenotype. In pharmacology, interaction occurs when combination of multiple drugs produce an effect different of what is expected of individual drugs due to chemical reactions. In these examples and others, interaction is related to the outcome variable of the experiment and its incidence is directly observed on pairs of objects under study. The identification and description of interaction in these experiments is made independent of the probability model used to analyze the experiment.

In contrast, interaction in statistical literature is not attached to the outcome variable but instead related to the design variables. In addition, interaction is conceptualized as a parameter whose value needs to be estimated from the data. Since interaction is assumed to be measured on a continuum, one needs to collect quantitative data from the experiment to estimate the interaction. An underlying model is also required in this setting to identify or describe the interaction. Often, the model used is either regression model or the two-way or higher-way ANOVA models. We say that two variables interact if the effect of each of these variables on the response variable depends on the value of the other input variable. More formally, a pair of variables is said to interact if the joint effect of the pair equals to the sum or product of marginal effects of the pair. This implies that the incidence of interaction can only be detected by comparing joint distribution of the pair with marginal distributions of variables in the pair.

A typical assumption of the classical models of interaction is that the outcome response variable is a continuous variable unless a discrete response variable has undergone a continuous transformation as in e.g. logistic regression or generalized linear model. This requirement typically limits the use of classical models in practice since researchers in all fields of science often face the situation where they have to measure some continuous variable of interest but for which they lack an accurate and reliable measuring instrument. One approach to overcome this difficulty is through methods of statistical quantification in which values of the unknown continuous variable are estimated based on assumptions relating the variable with certain observable quantities or events. An example is when discretizing the variable into several discernible categories or classes. More formally, suppose an experiment is conducted on pairs of objects with outcome variable a continuous variable measuring the interactions among the pairs, where interaction is defined as any quantitative relation between a pair of objects. Furthermore, suppose it is hard (or impossible) to measure the response variable numerically but easy to code its values into ordered, ternary categories such as low, moderate, and high levels of interaction. For instance, in the protein interaction example above, the response variable is the propensity of a pair of amino acids to form a motif sequence. However, since propensity is a variable which is hard to measure quantitatively but easy to observe its intensity level for a pair of amino acids, one can alternatively run the experiment with data collected on levels of propensity such as low, moderate, and high levels of propensity between pairs of amino acids. It might come as a surprise to know that this partial information on the response variable together with the information contained in the design of the experiment suffice to recover the actual interaction values and to estimate the correlations between pairs of objects in the experiment.

To intuitively see why this approach works in theory, consider an ensemble of *n* objects such that for each pair of objects the researcher makes a note on whether the interaction between the pair of objects is low, moderate, or high. We may think of the objects as *n* points in space, and of the interactions as angels between pairs of vectors emanating from origin to the points. The order of the $n(n-1)/2$ angels can be selected arbitrarily provided the *n* points are placed in a sufficiently large Euclidean space i.e. all possible arrangements of the angles are permissible in this space. (Alternatively. for arbitrary set of $n(n-1)/2$ positive numbers, it is always possible to find *n* points in a Euclidean space of dimension at least $n-1$ with pairwise distances the set of numbers). This implies a scale *w* and two non-negative constants *a* and *b* exist such that the interactions $w_{ij}$ between pairs of objects *i* and *j* on this scale are in such a way that all interactions labeled low are smaller than *a*, all interactions labeled moderate are between *a* and *b*, and all interactions labeled high are larger than *b*. Now, if variance of the interactions is maximized, or equivalently, if dimensionality of the fitted space is minimized, then with some further restrictions the solution turns out to be tight and often unique (for details and further discussion, see Shepard (1962)).

As an example of these types of problems consider a researcher in a drug industry experimenting with drug combinations to explore interactions between pairs of drugs. The researcher classifies a drug combination into one of three groups: synergistic group which is when the combination of drugs results in a smaller dosage of drugs to achieve the same level of response effect, antagonistic group which requires increasing the dosage of one or both drugs in the pair to achieve the same level of response effect, and finally the additive group which requires no significant changes in the dosages of either drug in the pair for the same response effect. Synergistic drugs are characterized, according to Loewe additivity model (Chou (1984)) to have Combination Index significantly below unity, antagonistic drugs have CI significantly above unity, and additive drugs have CI approximately equal to unity. Some researchers in the drug industry groups arbitrarily choose values .8 and 1.2 as their cutoff points, i.e. they choose *a*=.8 and *b*=1.2 for the groups bounds; others, choose these values experimentally.

In the next section, we use the framework of weighted graph theory to introduce the Quantification of Qualitative Responses (QQR) technique to quantify all pairwise interactions between *n* objects. The problem is then reformulated in terms of convex optimization problem with a positive semidefinte condition on the weights, known in the literature as Semi-Definite Programming problem. We utilize properties of SDP to derive solutions for the proposed QQR technique. In Sections 3, we explore properties of QQR technique and decompose the interactions into components. We also construct an ANOVA-like table to test significance of the interaction components and show how to interpret the output values. In Section 4, three applications of the technique are presented, one in biochemistry, a second in pharmacology, and a third in computer imaging. In these applications, we show how to apply the technique, choose the parameters values, derive solutions, and interpret the solutions. We also discuss the various properties of the technique through these applications including optimality of the solutions. In Section 5, we consider situations where parameters *a* and *b* are unknown and extend QQR technique to these general cases. In Section 6, we compare the QQR method with other methods of quantification and finally , in Section 7, we conclude with general remarks for future work on the topic.

2- The Technique
Consider a simple undirected complete weighted graph with *n* vertices or nodes such that an edge connects any two distinct nodes (i.e. except between a node with itself). Assume the edges of this graph are partitioned into three distinct classes denoted by $S^<$, $S^<>$, and $S^>$ where each of the two classes $S^<$ and $S^>$ includes at least one edge, but $S^<>$ may be empty. Denote an edge between nodes $i$ and $j$ ($i \neq j$) by the pair ($i,j$) and let $w_{ij}=w_{ji}$ be the weights assigned to edge ($i,j$) for all $i<j$ in such a way that edges in $S^<$ are assigned weights less than parameter $a$, edges in $S^>$ are assigned weights larger than parameter $b$, and edges in $S^<>$ are assigned weights between $a$ and $b$, for given parameters values $a$ and $b$ where $0<a\leq b$. Denote by $\delta_{ij} = \delta(i,j)$ the sign function on the edges defined as

$$\delta(i,j) = \begin{cases} -1 & \text{if } (i,j) \in S^< \\ 0 & \text{if } (i,j) \in S^{<>} \\ 1 & \text{if } (i,j) \in S^> \end{cases}$$

The problem is to choose the non-negative weights in such a way to maximally discriminate between the nodes in the two classes $S^<$ and $S^>$. Assume the weights are positioned in a square matrix $W$ whose rows and columns are indexed by the nodes, and where the weights are the off-diagonal entries. Clearly, $W$ is a symmetric matrix whose diagonal entries are not yet defined. If $W$ is of rank $k \leq n$ then there exists a set of $n$ $k$-vectors $v_i$ associated with the $n$ nodes such that the weight $w_{ij}$ of edge ($i,j$) is the inner product of its corresponding nodes vectors $v_i$ and $v_j$, i.e. $w_{ij}=v'_i v_j$. Now, if the diagonal entries of this matrix are defined to be the norms of the vectors associated with each node then $W$ is well defined and is positive semidefinite since it is the Grammian matrix of the nodes vectors. For technical reasons, the diagonal entries of $W$ will be left unspecified but assumed to be non-negative numbers whose sum is bounded by a constant.

In applications, the nodes are proteins, genes, drugs, compounds, or any other entities whose pairwise relations or interactions (denoted by the edges) are unknown. The weights $w_{ij}$ are assumed to be the interactions values, and also values for an unobservable continuous outcome variable $w$ underlying the interactions. To carry out the analysis, data on pairs of nodes are obtained by classifying each pair of nodes into a ternary, ordered scale such as the scale: no or weak correlation, moderate correlation, and strong correlation, or the scale: dissimilar, neither dissimilar/similar, and similar, or more generally the scale: low, moderate, and high levels of interaction. These categories are denoted by $S^<$, $S^{<>}$, and $S^>$, respective. At this point we are ready to formally define and solve the QQR problem.

**Theorem 1.** Given a complete weighted graph with $n$ nodes and with edges between all distinct nodes divided into three classes $S^<$, $S^{<>}$, and $S^>$, then there exist optimal weights $w_{ij}^*$ which discriminate maximally between the two non-empty classes $S^<$ and $S^>$. The optimal weights are the solutions of the following semidefinite programming problem:

$$\text{Max} \sum_{i<j} \delta_{ij} w_{ij}$$

Subject to:

For $i \neq j$, $\begin{cases} w_{ij} \geq b, & \delta(i,j)=1 \quad \text{when } (i,j) \in S^> \\ w_{ij} \leq a, & \delta(i,j)=-1 \quad \text{when } (i,j) \in S^< \\ a < w_{ij} < b, & \delta(i,j)=0 \quad \text{when } (i,j) \in S^{<>} \end{cases}$

$$\sum_i w_{ii} \leq nR$$
$$0 \leq w_{ij} \leq R, \quad \forall i,j$$
$$W \geq 0,$$

where $a$, $b$ are given non-negative constants, and $R$ is a suitable estimate of range of quantification.

Proof. A solution for QQR problem exists if and only if its corresponding SDP solution exists. Existence of solution for SDP problem follows from discussion on construction of *W*. The SDP problem is feasible provided the value of parameter *R* is suitably selected for given values of *a* and *b*. □

Remark. It is important to correctly specify the value of *R* In Theorem 1 in order to ensure optimality of the solution. When *R* is too small the quantification yields infeasible solutions, and when it is too large, the quantified values are restricted and thus some interactions are quantified at values *a* and/or *b*. In practice, one has to try several values for *R* before finding a good estimate that yields a feasible solution yet unrestricted in value. Such a solution is found in the border area between feasible and infeasible solutions. Notice that when *R* is near the border line between feasible and infeasible solutions the QQR solution is often very sensitive to the value of *R*. A slight change in the value of *R* in the vicinity of the border line results in dramatic changes in the QQR solution values. The difficulty of correctly specifying the value of *R* will be removed in Section 5, however, when parameters *a* and *b* are assumed unknown and the value of *R* is shown to be arbitrary.

3- Properties

Theorem 1 provides a procedure for deriving the solution of QQR problem based on semidefinite programming. There are a number of algorithms that derive solutions for semidefinite programming, and these algorithms enjoy many good properties including robust and efficient solutions in a straightforward and easy to use steps. In addition, semdefinite programming algorithms use interior point methods which have complexity order polynomial in time, Boyd (2004).

Consider once again the matrix *W* of interactions values between pairs of the *n* objects (or nodes) in the experiment and recall the interactions are not directly observable in our experiments; only their class labels are assumed to be observed. If the interaction space of all the pairwise interactions is of dimension *k* ≤ *n* then there exists a matrix *U* of dimensions *n*×*k* such that *W = UU'*. In other words, each interaction in the experiment can be decomposed into *k* components each of which is the product of two parameters corresponding to row and column effects i.e. each interaction is written in terms of the corresponding nodes effects. More specifically, the model in this paper assumes

$$w_{ij} = \lambda_1 u_{i1} u_{j1} + \lambda_2 u_{i2} u_{j2} + \lambda_3 u_{i3} u_{j3} + \cdots + \lambda_k u_{ik} u_{jk} + \varepsilon_{ij}$$

where $u_i = (u_{i1}, u_{i2}, \cdots u_{ik})'$ are a set of normalized and pairwise orthogonal vectors, $\lambda_i$ are scale parameters, and $\varepsilon_{ij}$ are i.i.d random variables with zero mean and constant variance $\sigma^2$. Clearly the model is nonparametric; however, occasionally the errors are also assumed to be normally distributed to derive tests of significance. Mandel (1969, 1971) treats the above interaction model similar to ANOVA model and partitions the sum of squares of interaction into *k* sums of squares corresponding to the *k* product terms in the model based on the following identity which can readily be shown:

$$\sum_i \sum_j w_{ij}^2 = \lambda_1^2 + \lambda_2^2 + \cdots + \lambda_k^2 .$$

Clearly, each squared lambda in this identity is associated uniquely with a product term in the model. Mandel also defines an analogue to the notion of degrees of freedom, one for each squared lambda. Consequently, a table similar to the ANOVA table is constructed for the terms of the model including tests of significance. In addition, he provides tables of simulated degrees of freedom of the first three sums of squares for various values of *n*. If we denote by $M_i$ the expected value of $\lambda_i^2$, and assuming the interactions are iid N(0, σ²) then Mandel (1969) shows that each of the ratios $\lambda_i^2/M_i$ is an independent estimate of error. In case the model includes only few interaction terms say, two or three terms, the rest are pooled to provide an estimate of error. The steps to derive the ANOVA table and the calculations involved are as follows..

Step 1. Solve the QQR problem using its SDP formulation in Theorem 1 and obtain the solution matrix *W*.

Step 2. Decompose *W* and obtain *n* eigenvalues $\lambda_1, \lambda_2, \cdots, \lambda_n$ and the corresponding eigenvectors.

Step 3. Select *k* where *k* is number of product terms retained in the model corresponding to largest *k* eigenvalues, (usually *k*=2 or *k*=3).

Step 4. Look up corresponding pseudo degrees of freedom $M_1, M_2, ..., M_k$ in Mandel (1969, 1971).

Step 5. Construct the following ANOVA table and test relevant hypotheses.

| Source | DF | SS | MS |
|--------|----|----|----|
| Total  | $n^2$ | $\sum_{i,j} w_{ij}^2$ | |
| 1st term | $M_1$ | $\lambda_1^2$ | $\lambda_1^2/M_1$ |
| ... | ... | ... | ... |
| *k*th term | $M_k$ | $\lambda_k^2$ | $\lambda_k^2/M_k$ |
| Error | $M_E$=subtract | $SS_E$=subtract | $MSE=SS_E/M_E$ |

In applications of QQR technique the question arises as to how interpret the interaction values $w_{ij}$. Because of the restriction *W* is positive semidefinite, *W* is interpreted as a covariance matrix and thus each off-diagonal $w_{ij}$ of *W* is the covariance value of the pair of nodes *i* and *j*. Moreover, because of the restrictions $w \leq R$ and $tr(W) = nR$, all the nodes have norms nearly equal to √R. As a result, *w/R* is the correlation between a pair of nodes, and it assumes only non-negative values. Since a correlation matrix with only non-negative entries is also a similarity matrix, *w/R* can also be considered a measure of similarity between a pair of nodes with value close to one indicating most similar nodes, and value close to zero indicating least similar nodes (or dissimilar nodes). Geometrically, each covariance value is proportional to the direction cosine of the angle between the pair of nodes vectors defining an interaction so that the maximum interaction occurs when the nodes of the two vectors are perfectly aligned, and the minimum occurs when they are perpendicular.

We conclude this section with an important property of QQR technique namely, the solution of QQR problem derived in Theorem 1 is in its most compact form meaning that it involves minimum number of parameters. In other words, the theorem shows that not only the QQR technique is optimal with respect to maximum discrimination criterion but with respect to the parsimony principle as well.

**Theorem 2**. Consider the QQR problem in Theorem 1 and let *W\** be an optimal solution i.e. a solution to its associated semidefinite programming problem. Then *W\** is also a solution of the following constrained rank minimization problem:

$$\text{Min } Rank(W)$$

Subject to:

$$\text{For } i \neq j, \begin{cases} w_{ij} \geq b & \text{if } (i,j) \in S^> \\ w_{ij} \leq a & \text{if } (i,j) \in S^< \\ a < w_{ij} < b & \text{if } (i,j) \in S^\diamond \end{cases}$$

$$\sum_i w_{ii} \leq nR$$

$$0 \leq w_{ij} \leq R, \quad \forall i,j$$

$$W \geq 0,$$

Proof. In Al-Ibrahim (2015) it is shown that maximizing $\sum_{i<j} w_{ij}$ for any positive semidefinite matrix *W* implies minimizing rank of *W* provided trace is bounded. Let $\tilde{W} = C \circ W$ where the circle denotes Hadamard product, and where the symmetric matrix *C* is such that for $i \neq j$, $c_{ij} = 1, -1, 0$ according to $c_{ij} \in S^>$, $S^<$ or $S^\diamond$, respectively, and with $c_{ii}$ equals to *n* for all *i*. To show *C* is positive semidefinite we first write $C = D + \sum_{i=1}^{n(n-1)/2} C_i$ where each $C_i$ is associated with an upper off-diagonal entry in *C*. $C_i$ is the zero matrix whenever the corresponding off-diagonal in *C* is zero, and when an off-diagonal in *C* is either 1 or -1 both corresponding diagonals equal to 1 and all remaining entries in $C_i$ are zero for all *i*. Clearly, each $C_i$ is positive semidefinite and thus it follows that their sum is also positive semidefinite. *D* is a diagonal matrix with diagonal entries $d_{ii} = n - c_{ii}*$, where $c_{ii}*$ are the degrees of the nodes which are non-negative; hence *C* is positive semidefinite, and by Schur Hadamard Theorem $\tilde{W}$ is also positive semidefinite. Next, notice that $trace(\tilde{W}) = \sum_i c_{ii} w_{ii} = n \, trace(W)$. We have thus shown that $\tilde{W}$ is positive semidefinite whenever *W* is, and that trace $\tilde{W}$ is bounded whenever trace *W* is. Consequently, maximizing $\sum_{i<j} c_{ij} w_{ij}$ implies minimizing rank of $\tilde{W}$. It remains to show that minimizing rank of $\tilde{W}$ implies minimizing rank of *W*. To this end, let *J* denote the *n×n* matrix with all entries equal to one, and notice that the largest eigen value of *J* is equal to *n* and all other eigen values equal to zero whereas the largest eigen value of *C* is larger than *n* and all other eigen values are non-negative. Thus, $C - J$ is positive semidefinite and thus $W \circ (C - J) = W \circ C - W$ is also positive semidefinite leading to conclusion that $rank(W \circ C) \geq rank(W)$. Hence, minimizing the left-hand side implies minimizing the right-hand side and this completes the proof of the theorem. □

4- Applications

In this section we borrow several examples from biology, pharmacology, and computer imaging in order to demonstrate the flexibility and wide range of applications of QQR technique and the various problems in science and engineering that can be solved with QQR technique. Other applications also exist but are not presented here for lack of space and because they are more fit in other disciplines. In all these applications, the data are analyzed with Matlab 7.3 using parser Yalmip (available online) and solver SDPT3. The computation times to carry out the calculations of these applications range from few seconds to a fraction of a minute.

4.1 First Application: Protein Interaction

We first consider an example from molecular biology to illustrate the utility and usefulness of the present approach to quantify interactions between entities in a system. More specifically, we will be interested in quantifying propensity for contact interactions between pairing residues across strands in β-barrel membrane protein. This will be based on quantification of protein. There are two broad approaches for protein quantification, total protein quantification and individual protein quantification. Total protein quantification methods comprise traditional methods such as the measurement of UV absorbance at 280 nm, Bicinchoninic acid (BCA) and Bradford assays, as well as alternative methods like Lowry or novel assays developed by commercial suppliers. Individual protein quantification methods, on the other hand, include enzyme-linked immunosorbent (ELISA) assay, western blot analysis, and more recently, mass spectrometry, among others.

In all the quantification methods mentioned above, observations are made and measurements collected on protein interactions. However, because of the complexity of these interactions, it is not clear how these measurements can be or should be used to capture propensity between pairs of residues in a protein. In particular, there is no unique or best way to quantify contact interaction between pairs of residues. It is, however, relatively easy to describe verbally the contact interaction between residues as strong, weak, or somehow strong/weak, based on certain observations, e.g. frequency of contacts. Jackups and Liang (2005) proposed a probabilistic model to quantify interactions contact between nineteen amino acid interstrands of β-barrel membrane protein residues. Their quantification aimed residues for different spatial locations and for interstrand pairwise contact interactions involving strong H-bonds, side-chain interactions, and weak H-bonds. In this section we quantify the contact interaction between the nineteen amino acid interstrands of β-barrel membrane protein residues of the weak H-bond based only on the classification data of the interactions. We will compare their quantifications with the QQR quantifications in this paper to see the connection between the two approaches.

In their study, Jackups and Liang (2005) defined eight distinct spatial regions for each transmembrane strand of β-barrel membrane proteins based on the vertical distance along the membrane normal and the orientation of the side-chain. After placing the origin of the reference frame at the midpoint of the membrane bilayer and taking the vertical axis perpendicular to the bilayer as the z-axis, they measured the vertical distance of each residue from the barrel along the z-axis. Interstrand pairwise interactions are then characterized with three different types of interactions, namely, strong regular hydrogen bonds between backbone N and O atoms, "non-H-bonded" side-chain interactions (interactions without strong backbone N-O hydrogen bonds), and weak $C_\alpha$-O hydrogen bonds between the $C_\alpha$ atom of one residue and a backbone oxygen on another strand. The strong N-O hydrogen bonds occur between residues on adjacent strands. The non-H-bonded interactions alternate with the strong hydrogen bonds along adjacent β-strands. The weak $C_\alpha$-O hydrogen bonds are displaced one residue along adjacent β-strands. Jackups and Liang (2005) calculated TransMembrane Strand Interaction Propensity (Tmsip) scales, which are the values of odds ratios comparing the observed frequency of interstrand contact interactions to the expected frequency. Their study was based on a very small dataset of known structures, comprising only 19 β-barrel membrane proteins.

Here, we use only thirteen amino acid residues out of the 19 protein samples since the data for the remaining six amino acid residues are incomplete. We follow the same classification as in Jackups and Liang (2005) and divide the interactions between all pairs of residues of the thirteen amino acids into three categories: 1) value less than 0.8, 2) between 0.8 and 1.2, and 3) larger than 1.2. The self-interaction propensity values are ignored since they do not fit into our framework. Jackups and Liang (2005) call the first category *spatial antimotif* and the last category *spatial motif*, provided the values are significant. Since the data are all ordered increasingly from smallest to largest, we coded all pairs in the first category to -1, all pairs in the second category to zero, and all pairs in the last category to one, in accordance with model specification described in Section 2. Self-interaction pairs are also coded to zero, but this should not be confused with codings of interactions in class $S^{<>}$, but rather the coding is meant to determine how the variances and covariances enter the objective function. Finally, the value of parameter $R$ was selected based on many trials in the border area of feasible and infeasible solutions to arrive at a solution that is feasible and also least restricted. The critical value $R=1.46755$ was found as best approximate value of $R$.

The results from this analysis are displayed in Table 1 and Figure 1. The table lists the input data to the QQR technique and also the output quantification values (the QQR values) for interactions of all pairs of the 13 amino acids. In Figure 1, the QQR values are plotted against the observed propensity values which are derived based on a probability model. The QQR values, on the other hand, are based on a nonparametric method which are also optimal in the sense of maximum variance and in the sense of minimum number of parameters. In other words, not only the QQR technique explains more between-group variance of the interactions with least number of parameters, but it does so for any probability distribution that satisfies the constraints. As for the interpretation of the QQR values, it was shown before that interaction is equivalent to correlation (or similarity) between a pair of nodes. Since the nodes here are the 13 amino acids and since the data are increasing levels of propensity to form motif sequences for pairs of amino acids, it follows that two amino acids with high propensity level (i.e. pair in category $S^{>}$) are highly correlated or very similar while two amino acids with low propensity (i.e. pair in category $S^{<}$) are not correlated or weakly correlated i.e. least similar or most dissimilar. We notice that while only classification data are needed to derive the QQR values, quantitative data in the form of observed frequencies of pairs of residues were required to derive the propensities. In summary, the QQR technique both costs less in assumptions and data requirement and offers more in terms of results and inferences with optimal properties.

**Table 1. Quantification with QQR technique (parameters $a=.8$, $b=1.2$, $R=1.46755$) using classification data in Jackups and Liang (2005) (Lower diagonals) classification data, -1($S^{<}$), 0($S^{<>}$), 1($S^{>}$). (Upper diagonals and diagonals) Quantified values using QQR technique**

| | Amino Acid | | | | | | | | | | | | |
|---|---|---|---|---|---|---|---|---|---|---|---|---|---|
| | A | C | D | E | G | H | I | K | L | M | N | Q | R |
| A | 1.468 | 1.017 | 0.800 | 0.668 | 0.858 | 0.973 | 1.127 | 1.200 | 1.082 | 0.816 | 0.973 | 0.800 | 0.800 |
| C | 0 | 1.230 | 0.937 | 0.971 | 0.993 | 0.943 | 1.098 | 0.911 | 0.949 | 0.897 | 0.942 | 1.035 | 0.934 |
| D | -1 | 0 | 1.468 | 0.517 | 0.830 | 0.800 | 0.840 | 1.200 | 0.788 | 0.269 | 0.800 | 1.200 | 0.800 |
| E | -1 | 0 | -1 | 1.463 | 1.209 | 0.801 | 1.120 | 0.410 | 0.837 | 1.219 | 0.799 | 0.834 | 0.846 |
| G | 0 | 0 | -1 | 1 | 1.461 | 0.424 | 0.861 | 0.814 | 1.122 | 0.705 | 0.423 | 0.709 | 0.815 |
| H | 0 | 0 | -1 | 0 | -1 | 1.468 | 1.352 | 0.800 | 0.531 | 1.158 | 1.468 | 1.200 | 0.800 |
| I | 0 | 0 | 0 | 0 | 0 | 1 | 1.453 | 0.898 | 0.851 | 1.274 | 1.351 | 1.165 | 0.917 |
| K | 1 | 0 | 1 | -1 | 0 | 0 | 0 | 1.468 | 0.813 | 0.311 | 0.800 | 0.800 | 0.429 |
| L | 0 | 0 | -1 | 0 | 0 | -1 | 0 | 0 | 1.398 | 0.623 | 0.531 | 0.809 | 1.098 |
| M | 0 | 0 | -1 | 1 | -1 | 1 | 1 | -1 | -1 | 1.459 | 1.158 | 0.870 | 0.853 |
| N | 0 | 0 | -1 | -1 | -1 | 1 | 1 | 0 | -1 | 0 | 1.468 | 1.200 | 0.800 |
| Q | 0 | 0 | 1 | 0 | -1 | 1 | 0 | -1 | 0 | 0 | 1 | 1.468 | 1.200 |
| R | 0 | 0 | -1 | 0 | 0 | -1 | 0 | -1 | 0 | 0 | -1 | 1 | 1.468 |

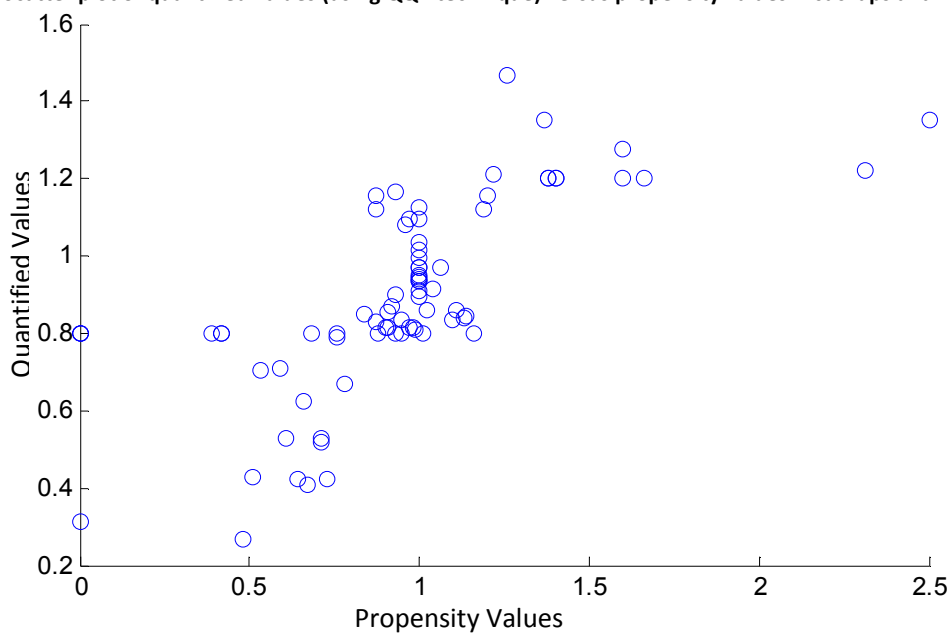

**Figure 1. Scatter plot of quantified values (using QQR technique) versus propensity values in Jackups and Liang (2005)**

It is customary to derive an estimate of error to assess the level of precision in quantification. Following the steps outlined before to construct the ANOVA table we first need to derive the eigenvalues from the matrix of quantified values in Table 1 and these equal to: 12.2557, 2.2642, 1.9398, 1.1648, 0.8023, 0.1914, 0.0924, 0.0175, 0.0085, 0.001, 0, 0, 0. The squares of these values are the sums of squares for the respective product terms in the interaction model, and they give a total sum of squares of 161.1382. The sum of squares for the first term is 150.2029, and this is about 93.21% of the total variation in the data. Hence, it is reasonable to assume the interaction model has only one product term and thus the sum of squares of error is the residual sum of squares which is 10.9353. Now, consulting the tables in Mandel (1969) for value of degrees of freedom for first product term (equals to approximately 35.39), this leaves us with 169 - 35.39 = 133.61 degrees of freedom for error and hence our estimate of error equals to .0818, i.e. a standard deviation of .2861. We can proceed to construct the ANOVA table and derive a test of significance for the interaction term, but this is unnecessary in this case.

4.2 Second Application: Drug Synergy

In the field of Biology there have been many studies focusing on interactions between specific drugs. There are three main types of interactions among multiple drugs: additive, synergistic and antagonistic. Drugs may not interact at all (called additive), may interact favorably (called synergistic), or may interact unfavorably (called antagonistic). Synergistic combinations are those with the greatest interest since they provide greater effect than would be predicted by simply adding together the effects of the components. In other words, synergistic drugs are drugs that are effective since they produce the same result an individual drug does but with less dosage.

There are two competing reference models of synergy, two methods that calculate the expected dose-response relationship for combination therapy as compared to mono-therapy: Loewe Additivity and Bliss Independence. Loewe additivity, which often is the preferred additive reference model, assumes that two inhibitors act on a target through a similar mechanism. To explore whether or not two inhibitors interact with each other, Loewe additivity uses the Combination Index defined as follows. Suppose we have the concentrations of two inhibitors ($[I_1]$, $[I_2]$) that individually achieve $X\%$ target inhibition. Then the concentration of inhibitors theoretically required to produce the same $X\%$ effect when used in combination ($[CI_1],[CI_2]$) is

$$CI = \frac{[CI_1]}{[I_1]} + \frac{[CI_2]}{[I_2]}$$

The *CI* compares the doses of inhibitors that experimentally produce the same level of inhibition individually and in combination. By finding the dose required for equal effect, we can determine whether the combination is effective at a lower total dose. An experimentally determined dose–response surface indicates synergy when its combination index is less than 1, additive when it equals to 1, and antagonistic when it is greater than 1, all within a margin of error. One advantage of Loewe model is that it predicts the combination of a drug with itself to be additive. The combination index is not the only way Loewe additivity is assessed. An alternative index transforms the scale so that zero indicates additivity whereas negative values of the index signify synergy and positive values signify antagonism. This index which is used by Cokol *et al* (2011) takes the form:

$$\alpha = log\left(\frac{x}{1-x}\right) - log(\frac{y}{1-y})$$

where *x* and *y* are drug normalized concentrations derived from the contours of the fixed target inhibition curve.

Cokol *et al* (2011) conducted large-scale drug synergy screens to predict synergistic interactions. The authors initially compiled a catalog of 113 known chemical/target relationships in *S. cerevisiae* from the literature, then integrated the resulting drug-target catalog with known synergistic genetic interactions. This yielded a set of 211 drug pairs predicted to be synergistic according to the parallel pathway inhibition model. The authors assessed these drugs according to parallel pathway inhibition model as well as according to bioavailability model. As part of their investigation, the authors hypothesized that different drugs have inherently different background synergy rates. They assessed the background synergy rates of 13 drugs that showed non-uniformity of background synergy rates. They concluded no single background rate of synergy can be used as a basis for comparison of success for a synergy prediction method. To account for the background rate of synergy, the authors considered the baseline synergy rate of each drug tested in 13×13 matrix, and tested whether drug pairs involving a given drug were more likely to exhibit synergy if they correspond to a genetic interaction. The test showed no evidence at the 5% significance level; hence concluded no correlation between genetic interaction and drug synergy. The authors also carried out network analysis in order to determine the smallest set of drugs that could "explain" all observed drug synergies. The 13 drugs with their first three letters abbreviation highlighted are listed below.

| **DYC**lonine | **FEN**propimorph | **HAL**operidol | **PEN**tamidine | **TAC**rolimus | **TER**binafine | **LAT**runculinB | **BEN**omyl | **STA**urosporine | **RAP**amycin | **TUN**icamycin | **CAL**yculinA | **BRO**mopyruvate |

In this section we aim to explore drug interaction through quantification. To carry out the analysis we only need to classify pairs of drugs into ordered classes representing some attribute related to level of interactions. Cokol *et al* (2011) studied the interactions between the 13 drugs and obtained experimental values for $\alpha$-Index. They used the values to classify all pairs of drugs into three classes:

synergistic if α<−.78, additive if −.78 ≤ α ≤ .68, and antagonistic if α>0.68. Here, we quantify the interactions between pairs of drugs based on combination index CI by coding the data into the following interaction classes: synergistic if CI<0.4584, additive if 0.4584≤ CI ≤1.9739, and antagonistic if CI>1.9739 where the boundaries here are selected by taking the exponents of the above intervals endpoints. As before, combinations in the lower class are coded to -1, those in middle class to zero, and those in upper class to one. The classification data are listed in the lower diagonals of Table 2 whereas the quantifications themselves are on and above the diagonal in the table. Also, in Figure 2, the quantified values are plotted against the synergy values computed by Cokol *et al* (2011) for comparison.

Table 2. Quantification with QQR technique (parameters *a*=0.458, *b*=1.974, *R*=3.679) using classification data in Cokol *et al* (2011). (Lower diagonals) classification data, -1($S^<$), 0($S^<>$), 1($S^>$). (Upper diagonals and diagonals) Quantified values using QQR technique

| | Drug | | | | | | | | | | | | |
|---|---|---|---|---|---|---|---|---|---|---|---|---|---|
| | DYC | FEN | HAL | PEN | TAC | TER | LAT | BEN | STA | RAP | TUN | CAL | BRO |
| DYC | 3.679 | 3.142 | 3.184 | 0.000 | 0.83 | 0.223 | 0.320 | 1.974 | 0.320 | 1.032 | 3.303 | 0.000 | 1.974 |
| FEN | 1 | 3.679 | 1.974 | 0.458 | 0.458 | 0.458 | 0.458 | 1.862 | 0.458 | 0.422 | 2.807 | 0.000 | 1.974 |
| HAL | 1 | 0 | 3.679 | 0.458 | 0.458 | 0.458 | 0.454 | 1.861 | 0.454 | 1.185 | 3.037 | 0.084 | 1.974 |
| PEN | -1 | -1 | -1 | 3.679 | 0.458 | 0.458 | 0.458 | 0.000 | 0.458 | 1.171 | 1.160 | 0.408 | 1.974 |
| TAC | 0 | -1 | -1 | -1 | 3.679 | 0.458 | 0.458 | 0.243 | 0.458 | 3.077 | 1.890 | 1.502 | 1.974 |
| TER | -1 | -1 | -1 | -1 | -1 | 3.679 | 0.458 | 0.622 | 0.458 | 0.000 | 0.759 | 0.250 | 1.974 |
| LAT | -1 | -1 | -1 | -1 | -1 | -1 | 3.679 | 3.192 | 3.679 | 1.974 | 0.813 | 3.458 | 1.974 |
| BEN | 1 | 0 | 0 | -1 | -1 | 0 | 1 | 3.679 | 3.192 | 1.613 | 1.974 | 2.685 | 2.292 |
| STA | -1 | -1 | -1 | -1 | -1 | -1 | 1 | 1 | 3.679 | 1.974 | 0.813 | 3.458 | 1.974 |
| RAP | 0 | -1 | 0 | 0 | 1 | -1 | 0 | 0 | 0 | 3.679 | 2.194 | 2.696 | 2.544 |
| TUN | 1 | 1 | 1 | 0 | 0 | 0 | 0 | 0 | 0 | 1 | 3.679 | .797 | 2.989 |
| CAL | -1 | -1 | -1 | -1 | 0 | -1 | 1 | 1 | 1 | 1 | 0 | 3.679 | 1.974 |
| BRO | 1 | 1 | 1 | 1 | 1 | 1 | 1 | 1 | 1 | 1 | 1 | 1 | 3.679 |

Figure 2. Scatter plot of quantified values (using QQR technique) versus synergy values in Cokol *et al* (2011)

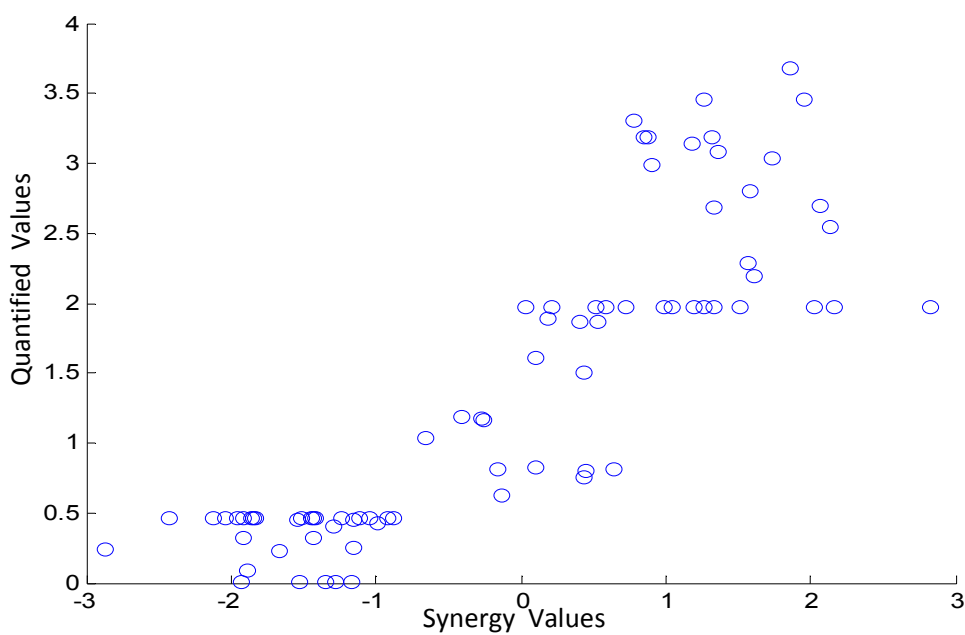

Notice from Table 2 and Figure 2 the accumulation of many quantifications at the boundary points $a^*=.458$ and $b^*=1.974$. This is often the case with QQR technique when the values of parameters $a$ and $b$ are assumed known. The reason for this phenomenon is because of lower and upper bounds on the interactions, but also because of poor choice of the parameters values. The accumulation of points still occurs at the boundary points when the parameters are unknown but with less frequency since they tend to spread out over the entire region $(0, R)$ and accumulate near the boundaries especially at lower boundary zero.

Because pairs of drugs that interact synergistically (i.e. favorably) have smaller CI values while those that interact antagonistically have large CI values, it follows that pairs of drugs with smaller interactions (or quantified values) imply better combinations. However, since interactions are equivalent to correlations, weakly correlated drugs are expected to be synergistic and highly correlated drugs are expected to be antagonistic. Geometrically, this implies pairs of drugs which are angularly large are expected to be synergistic while pairs of drugs which are angularly small are expected to be antagonistic. In Figure 3, a map is drawn for the 13 drugs with their defining vectors emanating from origin to the coordinates of the first two principal components. The figure gives a visual representation of the quantifications in Table 2. Notice, however, some distortion of this rule for certain pairs of drugs due to the fact that only the first two components are used in the figure and these account for only about 67.43% of the total variation in the quantifications.

One of the objectives in Cokol *et al* (2011) study was to find out if a smaller set of drugs could be found that explain the observed drug synergies, and the authors used network analysis to investigate this hypothesis. Another way to investigate the hypothesis is to perform principal components analysis, which involves deriving eigenvalues and eigenvectors of the quantification matrix. One has then to look at the first few principal components which account for most of the variation in the quantification values and look for comparisons and clusters of points in the data.

**Figure 3. A map of 13 drugs showing synergistic combinations and antagonistic combinations in terms of angular measure**

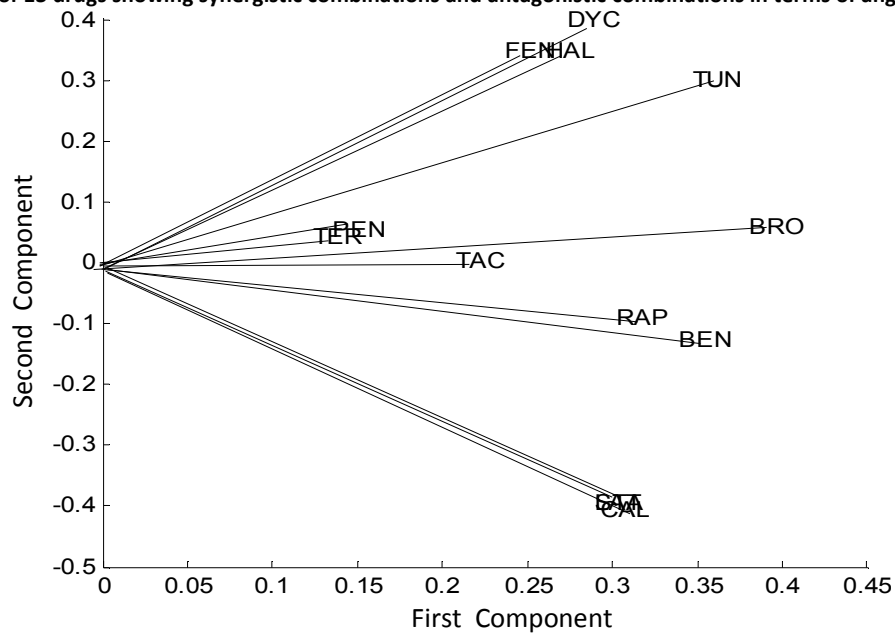

**Table 3. Eigenvalues and eigenvectors of matrix of quantifications in Table 2**

| | Eigenvectors | | | | | | | | | | | | | Eigen values |
|---|---|---|---|---|---|---|---|---|---|---|---|---|---|---|
| | DYC | FEN | HAL | PEN | TAC | TER | LAT | BEN | STA | RAP | TUN | CAL | BRO | |
| 1 | 0.267 | 0.238 | 0.256 | 0.125 | 0.205 | 0.111 | 0.297 | 0.348 | 0.297 | 0.315 | 0.341 | 0.291 | 0.373 | 21.72 |
| 2 | 0.410 | 0.353 | 0.348 | 0.025 | 0.004 | 0.040 | -0.374 | -0.115 | -0.374 | -0.130 | 0.321 | -0.411 | 0.060 | 10.53 |
| 3 | -0.192 | -0.217 | -0.146 | 0.367 | 0.561 | 0.099 | -0.212 | -0.385 | -0.212 | 0.389 | 0.089 | 0.006 | 0.197 | 6.11 |
| 4 | 0.172 | -0.028 | 0.067 | -0.415 | 0.309 | -0.710 | -0.045 | -0.005 | -0.045 | 0.298 | 0.063 | 0.110 | -0.290 | 4.43 |

Four components of the quantifications in Table 2 were derived and listed in Table 3. The first component has all its loadings positive and it captures the shared potency of the thirteen drugs i.e. their overall effect. It alone explains about 45.41% (=21.72/47.827 100%) of total variation in the quantified values. The other three components contrast some other properties of the drugs that are shared by only a subset of drugs but not the remaining drugs. The second component, for example, seems to be a contrast between the first six drugs and the next six drugs except the eleventh (with last drug almost excluded), which according to Cokol *et al* (2011) are, respectively, the *promiscuous* and *chaste* synergizers. In fact, these authors hypothesized and tested that promiscuity and chastity are intrinsic drug properties that would predict drug interaction behavior. One may consider the emergence of this component in Table 3 as an additional evidence to the validity of this hypothesis. Notice that this component accounts for about 21% of the total variation of the quantified values.

4.3 Third Application: Computer Imaging

In this application of QQR we take a twist at the notion of interaction, where here interaction is interpreted as a measure of similarity/dissimilarity between pairs of objects. Techniques that use measures of similarity/dissimilarity to draw maps are known in the literature as Multidimensional Scaling (MDS) methods with input data some measurements of similarity/dissimilarity. Here, we only need to qualify each pair of objects as either similar or dissimilar so that the data are binary classification of pairs of objects and not numeric measures of similarity/dissimilarity. Since our technique maps each input pair into one of class intervals $S^<$, $S^{<>}$, or $S^>$, and since the data are binary in this case, this implies that only two class intervals, $S^>$ and $S^<$, are in the range of the mapping but not the middle class interval $S^{<>}=[a,b]$. Accordingly, we must have $a=b$ in this application to allow for all positive numbers as possible values for quantification.

The primary objective of MDS methods is to draw a map of objects such that pairs of objects which are more similar are located closer to each other on the map than pairs of objects which are less similar (or more dissimilar). There are several versions of MDS, which differ in the distance metric used, type of transformation of input data, and assumptions on functional relations linking the input data to the distances. Classical MDS assumes linear link between the numeric measures of similarity/dissimilarity and the Euclidean distances in the map. On the other hand, non-metric MDS methods assume the data are only monotonically related to the unknown distances which, unlike metric MDS, could involve arbitrary distance metric. Within this context, QQR technique may be regarded as a binary MDS method with input data being classifications of pairs of objects either as similar or dissimilar objects. The output from QQR is a cross-product matrix, which when decomposed, produces a configuration of points in a map. In this map, closer points indicate the corresponding objects are similar, and distant points indicate they are dissimilar.

Metric and non-metric MDS methods have been successfully used in the literature of machine learning and related fields of artificial intelligence and pattern analysis to learn distance metrics for kernel based methods and other metric based methods. These techniques use different types of training data to uncover structure and learn embedding algorithms that map the data into lower dimensional maps. However, only recently have learning algorithms been developed to encode objects with binary measures of similarity/dissimilarity to produce points on a map. For example, in Li *et al* (2008) pairwise constraints are imposed on the objects to ensure generation of binary data that are used to learn a distance metric that maps the data into must-link group and cannot-link group which correspond to, respectively, class intervals $S^>$ and $S^<$ in our QQR technique. On the other hand, Mignon *et al* (2012) use these constraints to develop a new algorithm for learning distance metric which they call Pairwise Constrained Component Analysis. Still, Globerson *et al* (2007) take an approach, which is equivalent to our QQR approach, to learn an embedding algorithm that maps objects into a space of lower dimensionality such that similar pairs of objects are mapped closer on the map than dissimilar points. We will use one of the experimental data in Globerson *et al* (2007) to see how easily the solution can be recovered with our approach.

The data used by Globerson *et al* (2007) are images of a given object (e.g. airplane) taken at different azimuths (0, 20, 40, ..., 340), and at different elevations (30, 35, ...,70 degrees), for a given illumination level. The objective is to recover the manifold structure of azimuth and elevation from pairwise images. Intuitively, such a manifold is a cylinder with azimuth constituting the circle base of the cylinder and with elevations constituting the height of the cylinder (see Globerson *et al* (2007) for more details).

The data can be thought of as a two-way array with nine rows corresponding to the nine elevations, and with eighteen columns corresponding to the different azimuths levels under study. Clearly, there are a total of 162 pairwise images to assess. Following Globerson *et al* (2007), two images are considered similar if they differ by at most one level of azimuth and elevation. In other words, pairs of images which are either immediate row neighbors or immediate column neighbors in the above two-way array are considered similar while all other pairs of images are considered dissimilar except for images in the first and last columns which are also considered neighboring columns. To reduce computational load without distorting structure, we used only five elevation levels but kept all the eighteen azimuths levels. This resulted in 90 pairwise images in total. After coding pairs of images into zero on diagonals of an array consisting of all pairs of images, and coding an off-diagonal to one if a pair is similar, and -1 if a pair is dissimilar, we constructed the 90 by 90 similarity matrix which was subsequently analyzed with Matlab 7.3 installed on a regular desk laptop. We also used parser yalmip, which is available online, to assist in writing the code for Matlab.

Here, we present the results of QQR technique applied to above described data using parameters values *a*=*b*=.5, *R*=1. Figure 4 exhibits two plots of the first three principal components that were obtained by performing eigen-analysis on the solution matrix *W* (plot of third component vs. first component is not shown here since it is similar to first plot). The plots clearly show recovery of the cylindrical shape of the underlying structure. In the plots, the first component captures the elevation factor while the second and third components together capture the azimuth factor. Notice from first plot that elevation level one coincides with elevation level five, and similarly elevation level two coincides with elevation level four.

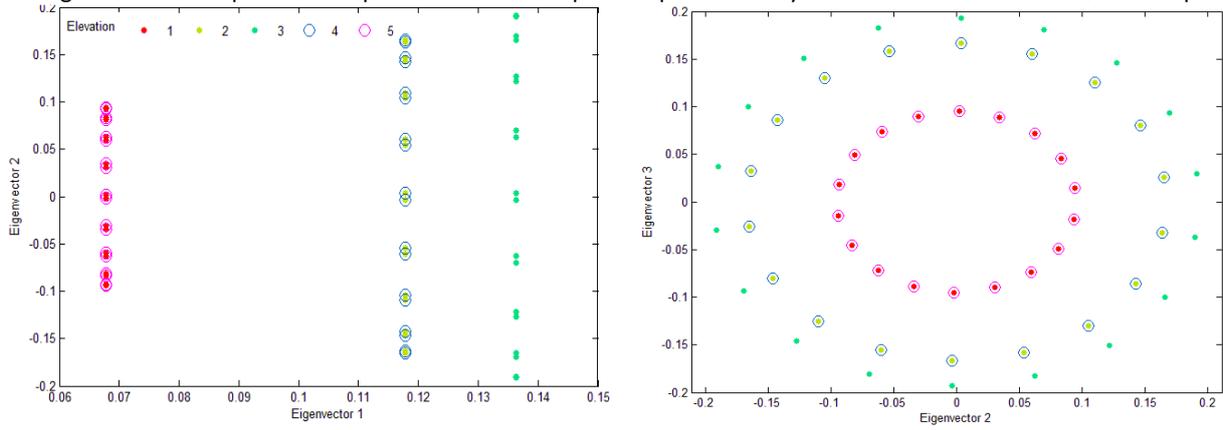

Figure 4. Plots of pairs of components from Principal Component Analysis of solution matrix of QQR technique

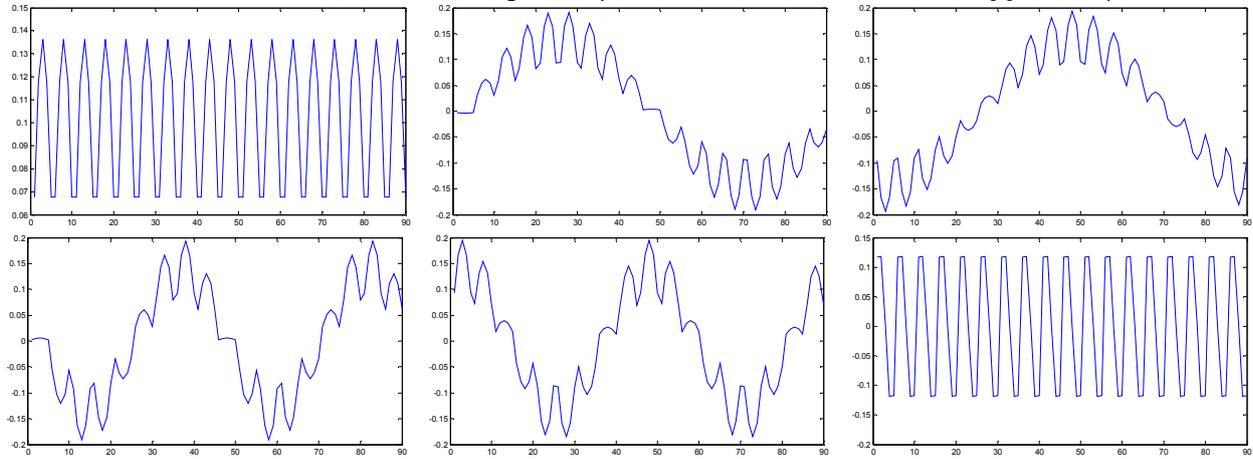

Figure 5. Plots of first six components from eigen-analysis of solution matrix derived from QQR technique

To further explore the nature of the solution, we plotted in Figure 5 the first six components of the solution matrix versus their serial order defined by combinations of levels of elevation and azimuth. These plots are remarkably similar to Fourier series plots of function *f(x)* over the period [-π, π] where the function here is that of cylinder with expansion $f(x) = a_0 + \sum_{n=1}^{\infty}\left(a_n \cos\frac{n\pi x}{L} + b_n \sin\frac{n\pi x}{L}\right)$, hence *L*=2π. Clearly, the first plot in the figure corresponds to the constant term $a_0$, the second plot to cos and the third to sin, for frequency n=1. The next two plots in the second row of Fig. 5 correspond to cos and sin terms for second frequency level n=2, and the pattern continues for higher-level frequencies if we plot more and more components. These plots show that QQR technique perfectly recovers structure in the data and, moreover, have easily interpretable dimensions for the solution.

5- Optimal Values for QQR Parameters

So far, we have assumed that values of parameters *a* and *b* of the quantifying intervals are known, and we had only to specify value of parameter *R* correctly to derive the solution. As noted from previous sections, the quantifications depend on values of *a* and *b* and they are not always easily predicted. We will now assume parameters *a* and *b* are unknown and seek to derive values for these parameters in an optimal way. Not only will this remove the arbitrariness of these parameters values, but also produces solutions with higher values for the objective function. The following theorem extends the QQR technique to the case where parameters *a* and *b* are unknown, and derives optimal values for these parameters together with value of parameter *R*.

**Theorem 3.** Consider the QQR problem stated and solved in Theorem 1, and assume the parameters *a, b* and *R* are unknown. Then, the optimal values for parameters *a* and *b* are

$$a^* = \underset{W \geq 0}{Min} \underset{(i,j) \in S^<}{Max} w_{ij} \qquad b^* = \underset{W \geq 0}{Max} \underset{(i,j) \in S^>}{Min} w_{ij}$$

and the value of parameter *R* is arbitrary.

Proof. From assumptions on QQR problem we have

$$a \geq \underset{(i,j) \in S^<}{Max} w_{ij}, \qquad b \leq \underset{(i,j) \in S^>}{Min} w_{ij}$$

Recall our criterion for optimality which is to maximally discriminate between the classes $S^<$ and $S^>$. Applying this criterion to the present case where *a* and *b* are unknown leads us to maximize *b* and minimize *a*; hence obtaining *a\** and *b\**. Equivalently, we may obtain values *a\** and *b\** by pretending an additional point has been included in $S^<$ corresponding to *a*, and an additional point has been included in $S^>$ corresponding to *b*, and then solve the QQR problem with *a* and *b* included in the objective function with appropriate signs. The solution of this problem consists of the triple (*W\**, *a\**, *b\**). For value of *R*, we need to show that if (*W\**, *a\**, *b\**) is a solution for QQR problem with value of parameter *R* equals to $R_0$ and with value of objective function $Q_0$ then ($\gamma W^*$, $\gamma a^*$, $\gamma b^*$) is also a solution for QQR problem with value of *R* equals to $\gamma R_0$ and with value of the objective function $\gamma Q_0$, for any $\gamma > 0$. To this end, we first notice that *W\**≥0 if and only if $\gamma W^* \geq 0$. Since all other constraints of QQR problem are linear in the variables then they equally hold for the transformed values. Also, because of the linearity of the objective function, its value equals to $\gamma Q_0$. This completes the proof of the theorem.

□

Although Theorem 3 derives optimal values for parameters *a* and *b*, yet in practice the value of *a\** always collapses to zero. The next theorem tells us why and also how to avoid this undesirable result.

**Theorem 4.** The QQR problem with unknown parameters values *a, b* and *R* always yields the degenerate solution *a\**=0 unless the parameters *a* and *b* are constrained by some restriction that excludes zero as a possible value of *a* such as the restriction *a*+*b*=*R*.

Proof. Denote by $A = \sum_{(i,j) \in S^<} w_{ij}$ and denote by $B = \sum_{(i,j) \in S^>} w_{ij}$, and notice that the objective is to maximize Obj= *B* - *A* + *b* -*a*. Since all the terms in this objective function are non-negative then clearly the maximum occurs when *A*=0 and *a*=0, provided a solution exists at these parameters values. The condition *A*=0 applies if and only if all the weights $w_{ij}$ in the first category $S^<$ are zero. Thus, we need to show the existence of a positive semidefinite matrix with off-diagonals zero for all entries in category $S^<$ which also satisfy the QQR problem constraints. We do this by construction. Let $L_{ij}$ be the *n×n* symmetric matrix with unity at its (*i,j*)th off-diagonal, and at *i*-th and *j*-th diagonals, zero otherwise. Then $L = \sum_{(i,j) \in S^>, S^>} L_{ij}$ is positive semidefinite since each $L_{ij}$ is positive semidefinite. If *D* is *n×n* diagonal matrix with non-negative values, and *W* is any *n×n* positive semidefinite matrix then $\tilde{W} = D + W \circ L$ is also positive semidefinite, where the circle here denotes Hadamard product. Notice that in $\tilde{W}$ all the off-diagonals corresponding to first category $S^<$ are zeros and hence $\widetilde{W}$ is a solution to the QQR problem with $A=0$ for any solution matrix $W$ of QQR problem. This completes the proof. ∎

Quantification with unknown parameters *a* and *b* can now be derived using the restriction introduced in Theorems 4, and this is illustrated next with the data in the previous sections. The quantifications from this procedure are then compared with those derived in previous sections when *a* and *b* were assumed known. To simplify the comparison, value of the scale parameter *R* is selected equal to *2* so that the mid range of the interval [*a. b*] is centered at unity, which is also the center of the interval for the case of known parameters values. The comparison is then made graphically by plotting the quantifications in the two cases as shown in Figures 6 & 7.

From Figure 6 one notes that, for interactions in first and last categories with values smaller than *a* or larger than *b*, the quantifications are almost always smaller when the parameters are unknown than when they are known (points in the figure fall below the line with slope one). Also, in Figure 7, the quantifications with unknown parameters result in several identical values whose corresponding quantifications are quiet different when the parameters are known e.g. several zero quantifications and several quantifications at value $a^*=.4584$. We note that quantification with unknown parameters often result in shorter intervals [*a,b*] compared to same intervals when the parameters are known. For example, in Figure 6, the intervals for known and unknown parameters quantifications were, respectively, [.8, 1.2] and [.835, 1.165] and in Figure 7, they were [.4584, 1.939] and [.7614, 1.6708]. On the other hand, the objective function always increases when the parameters are assumed unknown since we have more parameters to optimize. For Figure 6, the values of the objective functions were 4.5558 and 8.3332 for, respectively, known and unknown parameters quantifications, and for Figure 7, they were 117.6383 and 133.2051, respectively.

**Figure 6. Scatter plot of quantifications when parameters *a, b* are known and when they are unknown for data in Table 1**

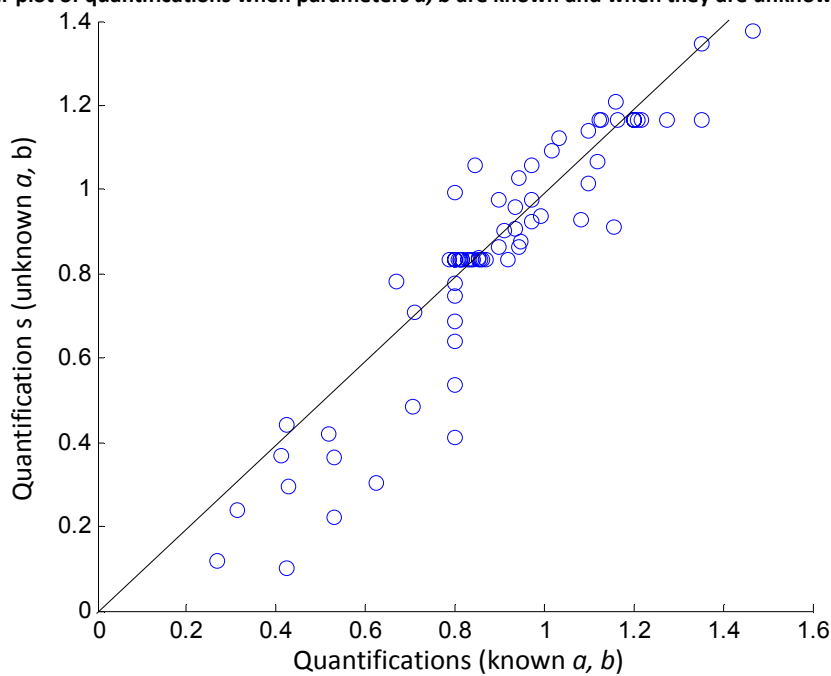

**Figure 7. Scatter plot of quantifications when parameters *a, b* are known and when they are unknown for data in Table 2**

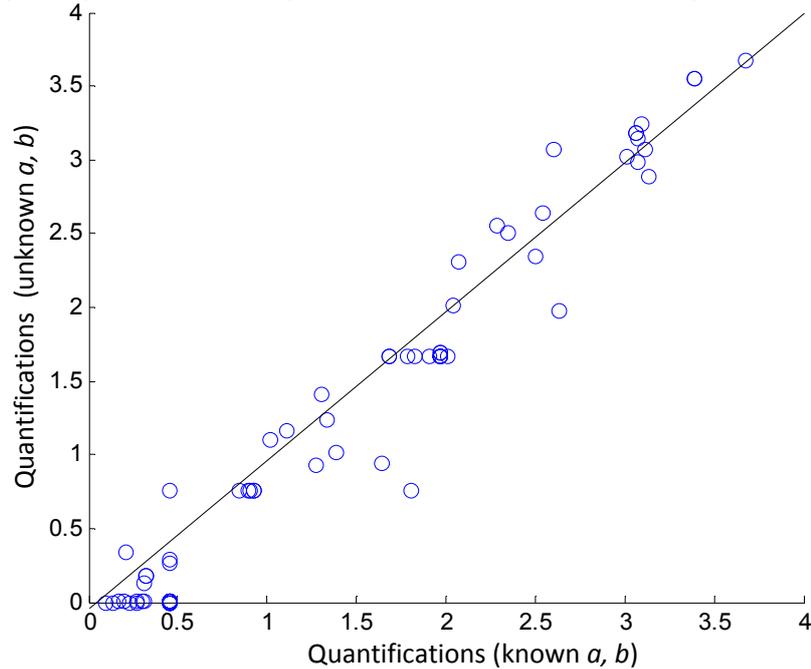

Finally, we applied QQR technique to data in third application with parameters values *a* and *b* unknown and with *R*=1. As expected, the optimal values for the unknown parameters converged to equal values $a^*=b^*=.39$ in the final solution although the equal value constraint was not imposed on the solution. The derived maps were quiet similar to maps obtained before with similar plots as in Figures 4 & 5 and therefore they are not shown here.

6-Comparison with Other Quantification Methods

In measurement theory, when the problem is to assess how much of an attribute each object in a set of objects possess, the method is traditionally referred to as a method of scaling. Perhaps the oldest method of scaling is the method known as the law of comparative judgment proposed by Thurstone (1927) in which he scaled a series of stimuli on a latent psychological continuum based on preference data on pairs of stimuli. Thurston's model was soon generalized to include any measurement model that generates scores from pairwise comparisons. A basic assumption of these models is that the underlying continuous variable follows a specified probability model so that the comparisons are made in terms of their probability values. In relation to QQR model, we note that QQR model is a more general model than the pairwise comparison model since it involves ternary classification of pairs of objects as opposed to binary classification for the paired comparison model. Furthermore, the QQR model is a nonparametric method which does not impose distributional restrictions on the responses. Yet, the most distinctive feature of the QQR model is that its parameters are estimated from data on a single run of the experiment whereas in the pairwise comparison model data from several runs of the experiment must be collected and aggregated since the technique analyzes frequencies. As a result, the QQR model can be used in a wider range of applications where data from only a single run of the experiment are available.

The pairwise comparison scaling method discussed above was extended in the literature in many directions and forms. Dual scaling and multidimensional scaling (MDS) are two extensions of the model developed under different conditions. Like the pairwise comparison method, dual scaling uses aggregate data in the analysis but without strict probability model requirements. On the other hand, classical MDS shares QQR method its unique feature of estimating the parameters from data on a single run of the experiment. Nevertheless, the data for this scaling method are assumed to be measured on a continuum, a requirement which greatly limits its use in practice. Another version of the MDS model known as non-metric MDS refines this requirement and assumes the input data to the analysis are the rank orders of the measurements. This is quiet similar to the QQR method which uses rank orders of three classes of pairs as input data to the analysis. Experiences with the two methods show that the QQR and the non-metric MDS methods often result in similar conclusions. In fact, a closer look at these methods reveals that both methods are based on same principles and same constraints on sets of pairs of objects. For more details on MDS method the reader is referred to Cox *et al* (2001), and for a good account on dual scaling method the reader is referred to Nishisato (1980).

Other related quantification methods also appear in the literature notably on the general multivariate categorical data analysis. These methods have long history of developments by many researchers, in many countries, using different approaches, and under different names. They appear in the sequel under the names: optimal scoring, optimal scaling, dual scaling, homogeneity analysis, multiple correspondence analysis, and scalogram analysis. They are based on different approaches to quantification including: the method of Reciprocal Averages by Horst (1935) and also by Fisher (1940), the ANOVA approach by Guttman (1941), the Principal Components Analysis approach by Burt (1950), and the Generalized Canonical Analysis approach by McKeon (1966). Tenenhaus *et al* (1985) show that all of these approaches lead to same equations and hence same solutions. Typically, the data for these approaches are multi-way contingency table and the quantifications aim to derive a lower dimensional common space for all the variables. The solution not only scales the different categories of all variables in a map, but also display the interdependencies of categories of different variables i.e. gives information on how the variables interact with one another.

Finally, we consider the popular factor analytic model which resembles QQR method in targeting the recovery of interactions (or correlations) of a latent factor with several manifest variables using the correlations among these observed variables as input to the analysis. The two methods, however, differ in their data requirements and model assumptions. Thus, although QQR and factor analytic methods share the objective of recovery of relationships among objects or variables, the two methods take completely different routes to this end.

7- Discussion and Conclusion
The basic idea in QQR technique is the familiar idea in homogeneity analysis in which between variances are maximized relative to total variance. This objective is achieved by maximizing the sum of quantifications in upper class $S^>$ as much as possible, and simultaneously minimizing the sum of quantifications in lower class $S^<$ as much as possible, for a fixed total variance. Similar to other homogeneity analysis techniques, maximizing variance results in dimensionality reduction. Thus, an equivalent way to phrase the objective of QQR quantification is that to reduce dimensionality of the problem as much as possible.

The QQR technique calls for classifying *a priori* interactions between all pairs of *n* objects into three ordered classes representing different levels of interaction. This requirement could be restrictive in some applications where prior information on certain interactions may be missing, or that we wish not to restrict the values of certain interactions to a given interval or class. The extension of our method to these more general designs is immediate and requires no further theory except that we include a fourth class of interaction where the parameters are unrestricted in the class i.e. consists of the union of all previous classes, hence the classes are no longer disjoint. An interaction is then classified into this new class only if there are reasons not to be included in any of the other classes. The analysis of this modified version of the technique is then proceeded in exactly the same way as before but with the additional class included in the SDP constraints and with constants $\delta_{ij} = 0$ in the SDP objective function.

The QQR technique was introduced in Section 2 for the case when the endpoints of the quantifying intervals are known, but it was extended in Section 5 to the case when they are unknown. And although the technique can be applied to both of these cases, it is recommended to always quantify with unknown endpoints since quantification with known endpoints values often result in biased solutions and generally dependent on values of the endpoints. QQR with known values for the endpoints should only be used when there is a natural choice for their values, or when there is some theoretical basis to backup the choice of values.

Finally, we emphasize that for the sake of smooth exposition of the material, the QQR technique was specialized in this paper to complete graphs. However, there is no reason to believe that such restriction is necessary or binding. One can extend the technique to many other graphs and in many other directions. Some possible directions for the extension of the technique includes: number of classes, type of classes (binary or multi-category), different classification rules, and of course different graphs or classification designs.